# Proiectarea şi implementarea unui portal HL7

Marius Cristian CERLINCĂ, Tudor Ioan CERLINCĂ, Cristina Elena TURCU,
Remus Cătălin PRODAN, Felicia Florentina GÎZĂ-BELCIUG

*Abstract*—This paper introduces some techniques used in developing and implementing an HL7 clinical data portal used in client-server architecture. The HL7 portal is used by non-HL7 applications that need medical data from HL7 servers. Also, the portal can translate a large number of HL7 terms between an indefinite number of languages.

## I. Introducere

HEALTH Level Seven (HL7) este o organizaţie non-profit implicată în dezvoltarea de standarde internaţionale legate de domeniul sănătăţii. Termenul HL7 este folosit de asemenea, ca referinţă la alte standarde create de această organizaţie: HL7 v2.x, v3.0, HL7 RIM, etc.

Se pot specifica standarde HL7 (versiunea 2.3.1): standarde conceptuale: RIM (Reference Information Model), standarde referitoare la documente: CDA (Clinical Document Architecture), standarde referitoare la aplicaţii: CCOW: (Clinical Context Object Workgroup), standard de schimb bidirecţional de mesaje: 2.3.x şi 3.0.

## II. Obiective. Prezentarea portalului HL7

### A. Obiective

Obiectivul principal al acestui modul este comunicarea respectând standardul HL7 între diverse aplicaţii şi module din cadrul unor sisteme informatice medicale. Astfel, acest modul este inclus (Fig. 1) în cadrul unui sistem informatic integrat pentru identificarea şi monitorizarea pacienţilor folosind tehnologii RFID, denumit SIMOPAC, care este prezentat în [1] şi [2].

Alte obiective specifice modulului sunt:
- să poată fi înglobat cu uşurinţă în cadrul altor module;
- să asigure funcţionarea corectă folosind sistemele de operare Windows 7/XP/2000, Linux şi Windows Mobile;
- să asigure confidenţialitatea datelor folosind elemente de autentificare şi criptare HL7 CCOW.

### B. Prezentarea portalului HL7

Portalul HL7 trebuie să asigure transferul de date medicale între diversele aplicaţii ale sistemului, chiar dacă acestea suportă sau nu standardul de mesagerie.

Modulele dezvoltate pentru portalul HL7 sunt:
- Modulul de comunicare HL7 V2/V3 care asigură comunicarea standardizată între diverse module din cadrul proiectului, precum şi cu alte aplicaţii medicale. Acesta va fi înglobat oriunde este necesară comunicarea datelor medicale în format HL7. Proiectarea şi implementarea modulului trebuie să asigure compatibilitatea cu următoarele sisteme de operare: Windows 7/XP/2000, Linux şi Windows Mobile. Va asigura datele medicale suplimentare celor de pe CIP, atunci când acestea sunt necesare;
- Criptare şi autentificare folosind HL7 CCOW, care asigură elementele de confidenţialitate privind datele medicale vehiculate în cadrul sistemului. Au fost implementate elementele de autentificare şi criptare HL7 CCOW. Modulul va asigura comunicarea criptată între modulele/aplicaţiile HL7 care suportă acest tip de comunicaţie.
- Managementul cheilor de criptare care păstrează şi distribuie cheile folosite în cadrul sistemului, asigură şi vehicularea acestora între diverse module. De asemenea, se asigură păstrarea şi distribuirea tipurilor de criptare şi autentificare folosite de fiecare sistem în parte.

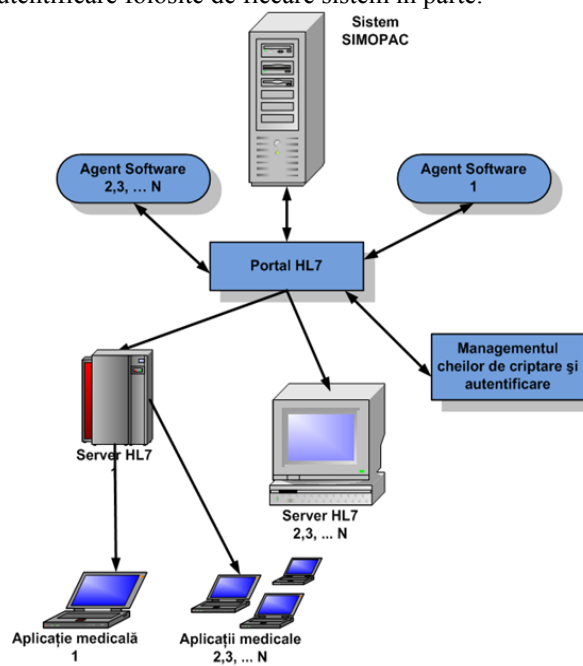

Fig. 1. Locul portalului HL7 în cadrul sistemului SIMOPAC.

### III. COMPONENTE ALE PORTALULUI HL7

Portalul HL7 oferă următoarele facilități:
- funcționarea corectă utilizând sistemele de operare Windows 7/XP/2000 și Linux;
- conectarea și autentificarea la servere HL7;
- obținerea de la aceste servere compatibile HL7 a datelor medicale referitoare la pacienți;
- transferul criptat al tuturor datelor vehiculate;
- interpretarea datelor din format HL7 într-un format cât mai apropiat de limbajul natural;
- traducerea, pe cât posibil, a datelor medicale obținute în limba română și nu numai;
- arhitectura sistemului oferă posibilitatea configurării unui set practic nelimitat de limbi străine, atâta timp cât acestea utilizează caracterele definite de standardul ASCII;
- asigură conectarea și autentificarea concomitentă a unui număr nelimitat de aplicații client în care sunt necesare date despre anumiți pacienți de la serverele de date medicale HL7;
- setul de comenzi suportate a fost proiectat astfel încât să ofere un suport cât mai complet în vederea obținerii de date medicale relevante;
- stocarea într-un fișier de evenimente, de exemplu, simopacServerInterpretare.log a tuturor conectărilor și comenzilor primite, respectiv a răspunsurilor date.

Pentru implementarea acestor facilități s-a stabilit arhitectura portalului HL7 (prezentată în Fig. 2), identificându-se modulele componente, care vor fi prezentate în continuare.

### IV. MODULUL DE CONECTARE ȘI TRANSFER AL DATELOR HL7

Modulul folosește mesageria HL7 V2 pentru transferul datelor clinice între un server extern HL7 și o aplicație SIMOPAC. Funcționarea acestui modul implică crearea unei conexiuni de tip socket TCP/IP, care se va conecta la un alt socket (adresă IP: port) server și va asigura vehicularea datelor clinice folosind modulul anterior pentru a asigura confidențialitatea datelor. Conexiunea tipică folosită este cea de tip HL7 LLP (Low Layer Protocol).

### V. MODULUL DE INTERPRETARE A DATELOR HL7

#### A. Facilități

Modulul de interpretare a datelor HL7 oferă următoarele facilități:
- permite interpretarea datelor clinice în limbi diferite;
- permite personalizarea mesajelor de interpretare;
- pot fi adăugate dinamic noi limbi care pot fi folosite pentru interpretarea datelor;
- execută comenzile primite de la clienți și întoarce datele clinice corespunzătoare dacă acestea există;
- permite conectarea unui număr nelimitat de clienți.

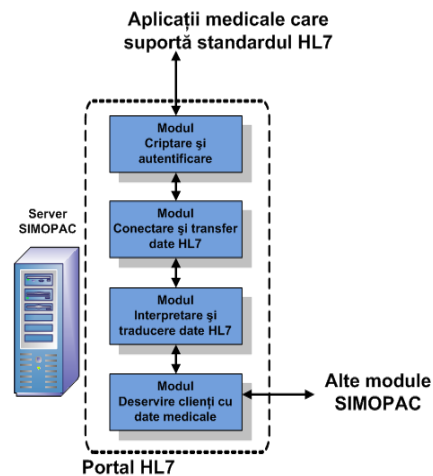

Fig. 2. Arhitectura portalului HL7.

#### B. Funcționare

Funcționarea acestui modul implică parcurgerea următorilor pași:
- citirea fișierului languages.txt care conține limbile străine suportate;
- citirea tuturor fișierelor folosite în interpretarea datelor în diferite limbi străine și inițializarea datelor aferente;
- crearea unui socket TCP/IP server unde se vor conecta eventualii clienți;
- se așteaptă conectarea clienților și se crează câte un socket de tip server pentru fiecare client în parte;
- răspunde mesajelor primite de la fiecare client și întoarce datele în limba setată de acesta;
- deconectează clientul la comandă și distruge perechea fir de execuție/ socket corespunzătoare.

#### C. Inițializarea datelor

Funcționarea sub-modulului este ilustrată în figurile următoare:

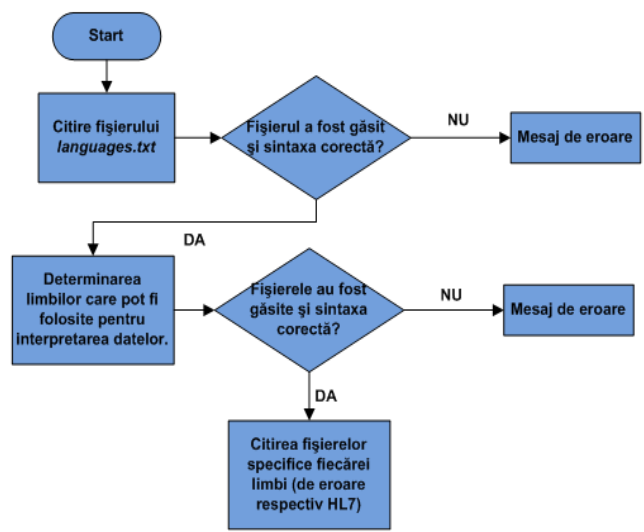

Fig. 3. Inițializarea modulului prin citirea fișierelor de limbi străine.

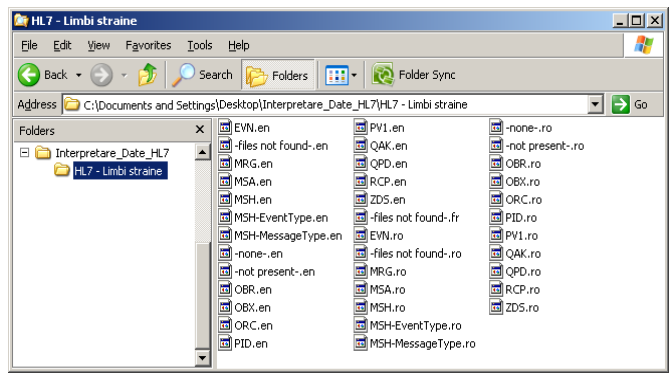

Fig. 4. Structura fişierelor de intrare folosite pentru interpretare.

*D. Fişiere folosite*

Fişierul languages.txt:

• este folosit pentru citirea limbilor diponibile pentru interpretare;

• conţine numele fiecărei limbi considerate, precum şi prescurtarea acesteia folosită pentru a identifica fişierele specifice. Exemplu:

Romana (ro)

English (en)

De exemplu, fişierele necesare interpretării datelor clinice în limba română sunt:

1) fişiere de eroare: -files not found-.ro, -none-.ro, -not present-.ro;

2) fişiere specifice comenzilor HL7 interpretate: EVN.ro, MRG.ro, MSA.ro, MSH.ro, MSH-EventType.ro, MSH-MessageType.ro, OBR.ro, OBX.ro, ORC.ro, PID.ro, PV1.ro, QAK.ro, QPD.ro, RCP.ro, ZDS.ro.

Fişierele necesare interpretării datelor clinice în limba engleză sunt:

1) fişiere de eroare: -files not found-.en, -none-.en, -not present-.en

2) fişiere specifice comenzilor HL7 interpretate: EVN.en, MRG.en, MSA.en, MSH.en, MSH-EventType.en, MSH-MessageType.en, OBR.en, OBX.en, ORC.en, PID.en, PV1.en, QAK.en, QPD.en, RCP.en, ZDS.en.

*E. Setul de comenzi al interpretorului*

Interpretarea datelor se va face la primirea de comenzi de la clienţii aplicaţiei. Acestea sunt prezentate în tabela 1.

Explicaţii comenzi şi valori returnate:

• conectare(IP, port, utilizator, parola): comandă trimisă de client pentru conectarea interpretorului la serverul unei aplicaţii HL7, întoarce OK în caz de reuşită sau NOK dacă conecarea nu a reuşit;

• utilizarePacient(CNP, limba): comandă care va seta pacientul curent; toate comenzile ulterioare primite de la clientul curent vor primi date referitoare la acest pacient, întoarce OK în caz de reuşită sau NOK dacă conecarea nu a reuşit;

• idExternPacient(): identificatorul extern asociat pacientului curent (extern referitor la aplicaţia HL7 interogată);

• idInternPacient(): identificatorul intern asociat pacientului curent (intern referitor la aplicaţia HL7 interogată);

• idAlternativPacient(): identificatorul alternativ asociat pacientului curent (alternativ referitor la aplicaţia HL7 interogată);

• nume(): numele pacientului curent (conectarea s-a realizat cu ajutorul codului numeric personal);

• numeFataMama(): numele de fată al mamei, util pentru profilul genetic al pacientului;

• dataNasterii(): data de naştere pentru pacientul curent;

• sex(): sexul pentru pacientul curent;

• rasa(): rasa pentru pacientul curent;

• adresa(): adresa curentă a pacientului;

• codulTarii(): codul ţării;

• numarTelefon(): numărul de telefon al pacientului curent;

• numarTelefonServici(): numărul de telefon al companiei angajatoare pentru pacientul curent;

• limbaNatala(): limba natală a pacientului curent;

• stareCivila(): starea civilă a pacientului;

• religie(): religia acestuia;

• numarContBancar(): numărul contului bancar pentru pacientul curent;

• codNumericPersonal(): CNP-ul pacientului curent;

• serieCarteIdentitate(): seria cărţii de identitate;

• minoritateaEtnica(): minoritatea etnică;

• loculNasterii(): locul naşterii;

• cetatenie(): cetăţenia;

• nationalitate(): naţionalitatea pacientului;

• ultimaEroare(): întoarce ultima eroare întâlnită.

TABELA. 1. STRUCTURA FIŞIERELOR DE INTRARE FOLOSITE PENTRU INTERPRETARE

| Comanda în limba română | Comanda în limba engleză |
|---|---|
| conectare(IP, port, utilizator, parola); | login(IP, port, user, password); |
| utilizarePacient(CNP, limba); | usePatient(SSN, language); |
| idExternPacient(); | getExternalID(); |
| idInternPacient(); | getInternalID(); |
| idAlternativPacient(); | getAlternateID(); |
| nume(); | getName(); |
| numeFataMama(); | getMotherMaidenName(); |
| dataNasterii(); | getDateOfBirth(); |
| sex(); | getSex(); |
| rasa(); | getRace(); |
| adresa(); | getAddress(); |
| codulTarii(); | getCountryCode(); |
| numarTelefon(); | getHomePhoneNumber(); |
| numarTelefonServici(); | getBusinessPhoneNumber(); |
| limbaNatala(); | getPrimaryLanguage(); |
| stareCivila(); | getMaritalStatus(); |
| religie(); | getReligion(); |
| numarContBancar(); | getAccountNumber(); |
| codNumericPersonal(); | getCNP(); |
| serieCarteIdentitate(); | getDriversLicenseNumber(); |
| minoritateaEtnica(); | getEthnicGroup(); |
| loculNasterii(); | getBirthPlace(); |
| cetatenie(); | getCitizenship(); |
| nationalitate(); | getNationality(); |
| ultimaEroare(); | getLastError(); |

*F. Adăugarea unei noi limbi pentru interpretarea datelor*

Pentru a adăuga o nouă limbă interpretorului trebuie parcurşi următorii paşi, considerându-se, de exemplu, limba franceză:

1) adăugarea unei noi linii în fişierul languages.txt (de ex: Francais (fr));
2) crearea fişierului -files not found-.fr cu conţinutul: "Dossiers HL7 non trouvés! Veuillez choisir une autre langue!";
3) crearea fişierului -none-.fr cu conţinutul: "Aucun";
4) crearea fişierului -not present-.fr cu conţinutul: "Pas présent";
5) traducerea conţinutului de tip medical al tuturor fişierelor specifice HL7 (EVN.en, MRG.en, MSA.en, MSH.en, MSH-EventType.en, MSH-MessageType.en, OBR.en, OBX.en, ORC.en, PID.en, PV1.en, QAK.en, QPD.en, RCP.en, ZDS.en) în limba franceză.

## VI. TESTAREA. STUDII DE CAZ

*A. Conectarea unui client la portalul HL7*

Pentru testarea modulului a fost dezvoltat şi un client generic care să execute comenzile descrise în secţiunea E. Portalul HL7 este executat pe platformă Linux, iar clientul pe platformă Windows:

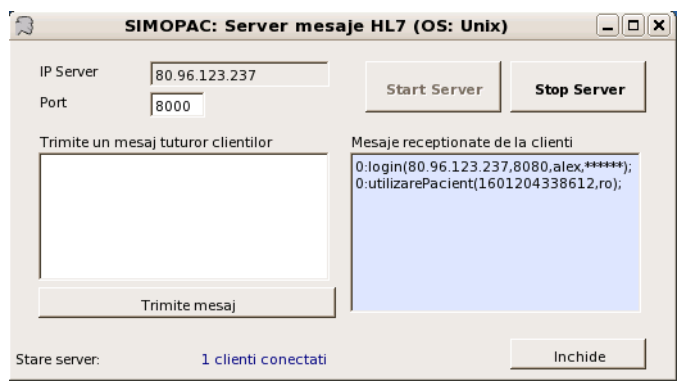

Fig. 5. Portalul HL7, funcţionare Linux.

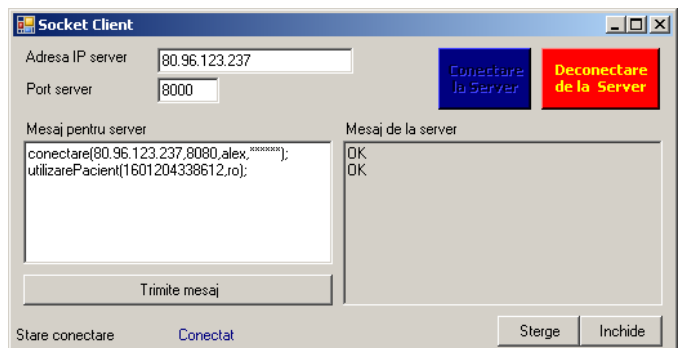

Fig. 6. Clientul: utilizarea comenzilor de conectare.

Observaţii:
• pacientul este unul imaginar, datele respectă condiţiile de confidenţialitate;
• şirul HL7 obţinut de la serverul HL7 a fost:

PID|||||C. Marius|Timpau|1975.09.16|M||Caucazian|Suceava, Jupiter Nr.1 Bl.121, Sc.B Ap.10|RO|0230420066|0720033743|RO|Necasatorit|Crestin Ortodox||1750916334996|||Roman|Suceava|||Romana||Romana|

• răspunsul "Nu exista date." Face referire la ultima comandă încheiată fără succes, în cazul nostru: serieCarteIdentitate();
• toate răspunsurile NOK se datorează faptului că datele respective nu există în şirul HL7 primit de la aplicaţia ţintă.

*B. Aplicaţii folosite în testare*

Pentru testarea portalului HL7 au fost folosite 3 aplicaţii compatibile cu standardul HL7:

- PatientOS;
- AccuMed EMR;
- serverul de mesagerie HL7 Mirth.

## VII. CONCLUZII. ELEMENTE DE NOUTATE

Elementele de originalitate ale portalului HL7 sunt:

1. Traducerea în premieră a unei părţi din mesageria HL7 în limba română.

2. Oferirea posibilităţii de interpretare şi parţial traducere a datelor din segmentele HL7 în orice limbă.

3. Oferirea unui mecanism simplu de adăugat noi limbi străine pentru interpretarea datelor.

4. Oferirea unei modalităţi ca în cadrul aplicaţiilor non-HL7 să se obţină şi procesa date aflate iniţial în format HL7.